\documentclass[twocolumn,prb,showpacs,preprintnumbers,amsmath,amssymb]{revtex4}

\usepackage{graphicx}
\usepackage{dcolumn}
\usepackage{bm}
\usepackage{epsfig}

\begin{document}


\title{An atomic mechanism for the boson peak in metallic glasses}

\author{U. Buchenau}
\email{buchenau-juelich@t-online.de}
\author{H. R. Schober}
\affiliation{%
Institut f\"ur Festk\"orperforschung, Forschungszentrum J\"ulich\\
Postfach 1913, D--52425 J\"ulich, Federal Republic of Germany}%
\date{March 14, 2008}

\begin{abstract}
The boson peak in metallic glasses is modeled in terms of local structural shear rearrangements. Using Eshelby's solution of the corresponding elasticity theory problem (J. D. Eshelby, Proc. Roy. Soc. {\bf A241}, 376 (1957)), one can calculate the saddle point energy of such a structural rearrangement. The neighbourhood of the saddle point gives rise to soft resonant vibrational modes. One can calculate their density, their kinetic energy, their fourth order potential term and their coupling to longitudinal and transverse sound waves.
\end{abstract}


\maketitle

\section{Introduction}

As yet there is no generally accepted explanation of the boson peak in the neutron or 
Raman scattering intensities of glasses. The boson peak is a broad peak at an energy 
transfer of a few meV, where simple crystals show only sound waves. Glasses seem to 
have a sizable excess of vibrations at this boson peak. At present, there is no 
agreement which forces drive these extra vibrations into the low-frequency region, 
though several possible explanations have been proposed
\cite{gurevich,schirmacher,elliott,nakayama,gotze,sokolov,parisi,ruocco}. 
Another controversial question \cite{ruffle,nizr} is whether the interaction with 
these vibrations is the physical reason for the Ioffe-Regel limit, the reduction
of the mean free path down to the wavelength of sound waves in the THz range.

The present paper proposes a detailed atomic model for the boson peak modes in 
glasses consisting of close-packed atoms or spherical molecules: The soft modes 
are ascribed to small regions with a pronounced shear misfit with respect to the 
surrounding matrix. In the following section II we describe the picture and derive the 
properties of these modes. Section III treats the connection to the extended soft 
potential model \cite{gurevich} and compares the predicted sound wave scattering and softening to 
experiment. Section IV discusses and summarizes the results.  

\section{The gliding-triangle mechanism}

\subsection{Shear strain defects}

\begin{figure}[b]
\hspace{-0cm} \vspace{0cm} \epsfig{file=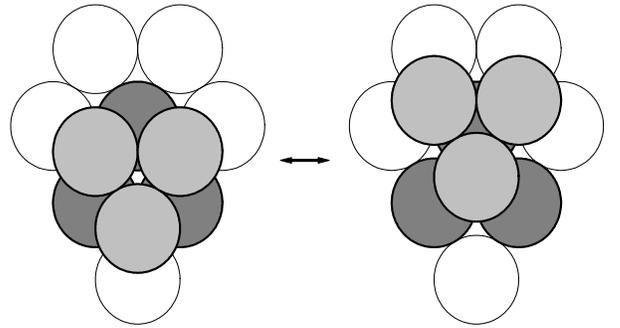,width=8cm,angle=0} \vspace{0cm}\caption{The gliding-triangle rearrangement of six closely packed spherical atoms.}
\end{figure}

The central concept of this model is a structural rearrangement of a limited region in the sample which changes its shape to a sheared one. In three-dimensional close packing, the lowest-energy structural rearrangement can be viewed as a gliding of a triangle of three close-packed atoms over an underlying close-packed plane (Fig. 1). On the left side of Fig. 1, the center of the upper triangle lies exactly over the center of a lower triangle within the close-packed plane. The central points of the six atoms form an octahedron. Gliding the triangle into the vertical direction in Fig. 1, one reaches again a stable position after a distance of $2r/\sqrt{3}$ ($r$ sphere radius), the situation at the right side of Fig. 1. The six atoms there form a pair of edge-sharing tetrahedra. Since the distance between close-packed planes is $2r\sqrt{2/3}$, the shear angle is $1/\sqrt{2}$ in radian, an angle of 40.5 degrees.

The physical problem of a small piece of matter able to transform to a sheared shape within an elastic matrix has been treated fifty years ago by several authors, notably by J. D. Eshelby \cite{eshelby,mura}. Here, we translate Eshelby's result into the usual convention, in which the shear angle $e$ and the shear stress $\sigma$ are related by $\sigma=G e$ ($G$ infinite frequency shear modulus). Let $v_i$ be the volume of the spherical inclusion and $e_i$ the shear angle difference between its two stable configurations (in the example of Fig. 1, $e_i=1/\sqrt{2}$ and $v_i=6v$, where $v$ is the atomic volume). For strictly harmonic potentials, the symmetric case of two equally strained configurations has the energy
\begin{equation}\label{ea}
	E_{a}=\frac{\gamma}{8} Gv_ie_i^2.
\end{equation}
The coefficient $\gamma$ is given by
\begin{equation}\label{gamma}
\gamma=\frac{7-5\sigma_P}{15(1-\sigma_P)},	
\end{equation}
where $\sigma_P$ is Poisson's ratio. Since Poisson's ratio lies between 0.1 and 0.44 for the known glasses, $\gamma$ lies between 0.48 and 0.57, close to 1/2.

Eshelby's solution divides the energy into two almost equal parts, one located in the inclusion and one outside. Their ratio is $\gamma/(1-\gamma)$. In the symmetric case, the inclusion would have to distort by $e_i/2$ to fit exactly into the unstrained hole in the surrounding matrix. From a structural point of view, this is the saddle point between the two stable configurations.

\begin{figure}[b]
\hspace{-0cm} \vspace{0cm} \epsfig{file=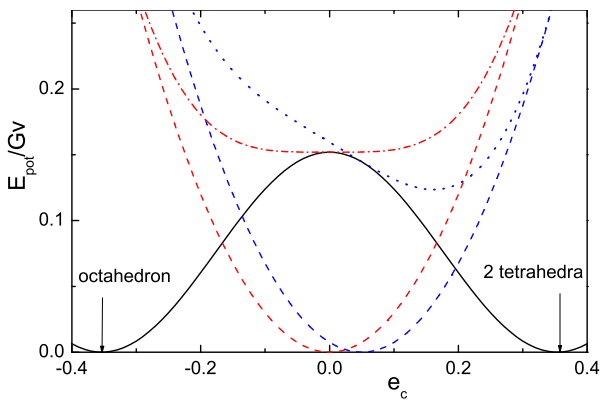,width=8cm,angle=0} \vspace{0cm}\caption{Potential energies for inclusion and matrix as a function of the inclusion shear $e_c$ around the structural saddle point, where the matrix equilibrium shear $e_m$ lies in the middle between the octahedron and the two tetrahedra of Fig. 1. The continuous black line is the inclusion energy, the dashed red line and the dashed blue line show the matrix energy for $e_m=0$ and 0.05, respectively. The red dot-dashed and blue dotted lines show the corresponding total energies.}
\end{figure}

In strictly harmonic approximation, Eq.~(\ref{ea}) holds and the structural saddle 
point indeed is a saddle point of the energy landscape. However, the harmonic 
approximation does not account for the flattening at the saddle point, at least as 
regards the distortion energy $E_c$ of the inclusion. For the inclusion, a better
description is by a cosine law 
\begin{equation}\label{cos}
E_c=\frac{3Gv}{4\pi^2}(1+\cos{2\sqrt{2}\pi e_c})
\end{equation}
shown in Fig. 2, where $e_c$ is the shear angle of the central inclusion, counted from the saddle point of the energy of the inclusion. This cosine law has the correct second derivative of $6Gv$ at $e_c=\pm 1/2\sqrt{2}$, the two structural energy minima of Fig. 1.

The embedding energy outside the inclusion depends on the difference between $e_c$ and the zero point $e_m$ of the shear of the matrix. In terms of $e_c$, the shear state of the hole after removing the inclusion and relaxing the matrix is $e_c=e_m$. Let us assume $\gamma=1/2$. Then the second derivative of this outside embedding term must also be $6Gv$. The total elastic energy $E_{el}$ is given by the sum of inside and outside contributions
\begin{equation}\label{etot}
E_{el}=\frac{3Gv}{4\pi^2}(1+\cos{2\sqrt{2}\pi e_c})+3Gv(e_m-e_c)^2.
\end{equation}

Note that the symmetric case at $e_m=0$ (the red lines in Fig. 2) has its minimum at $e_c=0$, with all the energy inside the inclusion and a vanishing restoring force constant for a displacement of $e_c$. Thus one does not get a saddle point of the energy landscape, but one gets a shear mode with zero restoring force constant at the structural saddle point. In this case, there is no elastic energy outside the inclusion and the creation energy $E_s$ for the soft mode configuration is
\begin{equation}\label{esoft}
E_s=\frac{3Gv}{2\pi^2}.
\end{equation}

In order to know the probability of finding such a soft gliding-triangle mode in a closed packed glass, one needs to know the ratio $Gv/k_BT_g$, where $T_g$ is the glass transition temperature, at which the glass falls out of the thermal equilibrium. For metallic glasses, this information can be extracted from a recent data collection \cite{samwer}. Fig. 3 plots the product $Gv$ for 30 metallic glasses at the glass transition as a function of $T_g$. There is a marked scatter, but on average $Gv/k_BT_g=17.6$ (the line in Fig. 3). With eq. (\ref{esoft}), this implies a ratio $E_s/k_BT_g$ of 2.67, i.e. a Boltzmann factor of 0.069. Thus one has to reckon with a sizeable number of nearly unstable resonant shear modes in a metallic glass. 

\begin{figure}[b]
\hspace{-0cm} \vspace{0cm} \epsfig{file=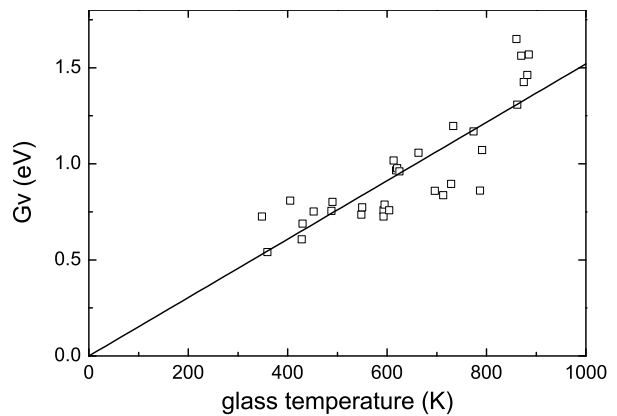,width=8cm,angle=0} \vspace{0cm}\caption{Energy $Gv$ as a function of the glass transition temperature $T_g$ in metallic glasses \cite{samwer}. The line depicts the average $Gv/k_BT_g=17.6$.}
\end{figure}

\subsection{Properties of soft shear modes}

The equilibrium value $e_{c0}$ of the inner coordinate $e_c$ for a given $e_{m0}$ 
is determined setting the first derivative of Eq. (\ref{etot}) with respect to $e_c$ 
equal to zero
\begin{equation}\label{eq}
e_{m0}=e_{c0}-\frac{\sin{2\sqrt{2}\pi e_{c0}}}{2\sqrt{2}\pi}.	
\end{equation}

The second derivative of Eq. (\ref{etot}) at $e_{c0}$ gives the restoring force constant
\begin{equation}\label{de}
	D_e=\frac{\partial^2E_s}{\partial e_c^2}=6Gv(1-\cos{2\sqrt{2}\pi e_{c0}}),
\end{equation}
which vanishes for $e_{m0}=e_{c0}=0$.

To calculate the kinetic energy of the mode, consider a sphere with a volume of $6v$,
corresponding to the six atoms in Fig.~1. 
Its radius $r_0$ is given by
\begin{equation}
	\frac{4\pi r_0^3}{3}=6v.
\end{equation}

Inside the sphere, the motion is a pure shear with amplitude $e=e_c-e_{c0}$, 
say $u_x=ey/2$, $u_y=ex/2$ and $u_z=0$. The motion of the outside has an amplitude 
decaying with the square of the distance from the center, which joins the inner 
shear amplitude continuously at the sphere boundary. The kinetic energy inside the 
sphere is
\begin{equation}\label{ekinin}
E_{kin,i}=
\frac{\rho\dot{e_c}^2}{8}\int_0^\pi\int_0^{2\pi}\int_0^{r_0}\sin^3\theta\ 
d\theta d\phi r^4dr,
\end{equation}
where $\rho$ is the density. The kinetic energy outside the sphere, 
in which the $r^4$ of Eq. (\ref{ekin}) is replaced by $r_0^6/r^2$ and $r$ is taken 
from $r_0$ to infinity, is a factor of five higher than the one inside. 
Replacing the density $\rho=M/v$, where $M$ is the atomic mass, one finally arrives at
\begin{equation}\label{ekin}
E_{kin}=\frac{1}{2}\ \frac{2^{1/3}3^{10/3}}{5\pi^{2/3}}Mv^{2/3}\dot{e_c}^2.
\end{equation}

For many purposes, it is useful to define a soft mode coordinate $A$ with the normal 
coordinate property $E_{kin}=\dot{A}^2/2$. For the soft shear modes
\begin{equation}\label{normal}
	A=\frac{2^{1/6}3^{5/3}}{5^{1/2}\pi^{1/3}}M^{1/2}v^{1/3}e_c.
\end{equation}

Consider the shear mode at the structural energy minima, say the octahedron of Fig. 1 at $e_m=-1/2\sqrt{2}$. Then, $D_e=12Gv$, and the mode frequency attains its maximum $\omega_{max}$ with
\begin{equation}\label{omax}
	\frac{2^{1/3}3^{10/3}}{5\pi^{2/3}}Mv^{2/3}\omega_{max}^2=12Gv.
\end{equation}

It is interesting to compare this frequency with the Debye frequency $\omega_D$, determined essentially by the transverse sound modes. The transverse sound velocity $v_t$ is given by
\begin{equation}\label{vt}
	v_t^2=\frac{G}{\rho}=\frac{Gv}{M}.
\end{equation}

The Debye sound velocity $v_D$ is given by the average
\begin{equation}\label{vd}
	\frac{3}{v_D^3}=\frac{1}{v_l^3}+\frac{2}{v_t^3},
\end{equation}
where $v_l$ is the longitudinal sound velocity, on the average for the metallic glasses \cite{samwer} a factor of 2.26 higher than the transverse one. For this case, $v_D=1.13v_t$. The Debye wave vector is $k_D=(6\pi^2/v)^{1/3}$, so on average
\begin{equation}\label{omd}
\omega_D=4.4\frac{v_t}{v^{1/3}}=2.72\ \omega_{max}.
\end{equation}
This shows that the gliding-triangle mode considered here has a maximum frequency of less than half the Debye frequency in the absence of any structural strain.

\subsection{Density of gliding-triangle modes}

In order to estimate the density of soft shear modes, one first needs the number of 
octahedra per atom in the disordered structure. In the crystalline close-packed 
structures, both fcc and hcp, one has one octahedron per atom. Though this number might 
easily be smaller in the disordered case, we will take it there also as one.

For each octahedron, there are twelve possibilities to distort to a pair of edge-sharing 
tetrahedra. There are four choices of the two triangles which are to glide against each 
other and for each pair there are three glide directions, $120^\circ$ rotated to each 
other. Consequently, there are two stable minima per atom, the octahedron and the 
edge-sharing pair of tetrahedra, and twelve saddle points between them. As shown in the Appendix, the twelve saddle points lie pairwise rather close to each other, with an angle of only 38.9 degrees between the two members of each pair.

At the glass temperature $T_g$, the saddle points are already sufficiently high in 
energy to be, in first order, negligible for the partition function, which can be 
approximated by a 
harmonic potential around the octahedron energy minimum at $e_{oct}=-1/2\sqrt{2}$. 
Close to the minimum, Eq. (\ref{eq}) gives $e_{c0}-e_{oct}=(e_{m0}-e_{oct})/2$. 
At the energy minimum, Eq. (\ref{etot}) takes the simple quadratic form
\begin{equation}\label{eharm}
E_{el}=\frac{3Gv}{2}(e_{m0}-e_{oct})^2,
\end{equation}
with the partition function
\begin{equation}\label{z}
Z=\sqrt{\frac{2\pi k_BT_g}{3GV}}	
\end{equation}
at the glass temperature $T_g$.

The probability $p_s=p(e_m=0)$ to find $e_m$ close to the structural saddle point is
\begin{equation}\label{ps}
p_s\approx\frac{3\exp(-E_s/k_BT_g)}{Z}=\sqrt{\frac{54Gv}{\pi k_BT_g}}\exp\left(\frac{-3Gv}{2\pi^2k_BT}\right).
\end{equation}
The factor 3 stems from the fact that there are three saddle point pairs per minimum, as explained in detail in the Appendix.

It is useful to introduce the coordinate $x$
\begin{equation}\label{x}
x=\sqrt{2}\pi e_{c0}
\end{equation}
which is zero at the structural saddle point and $\pm \pi/2$ at the two energy minima. From equs. (\ref{de}) and (\ref{omax}), one obtains for the frequency of the shear mode
\begin{equation}\label{om}
\omega=\omega_{max}\sin{x}.
\end{equation}

As one moves away in $e_m$ from the structural saddle point, the probability density changes only slowly, but $x$ and the restoring force constant of the mode change rapidly (see Fig. 2). The probability density of $x$ is related to the one of $e_m$ by the derivative $\partial e_m/\partial x$ calculated from eq. (\ref{eq})
\begin{equation}\label{demdx}
\frac{\partial e_m}{\partial x}=\frac{\sqrt{2}}{\pi}\sin^2{x}.
\end{equation}

The vibrational density of states of the soft shear modes is given by
\begin{equation}
g^\circ_s(\omega)=
\frac{p_s}{3}\frac{\partial e_{m0}}{\partial x}\frac{\partial x}{\partial \omega}=
\frac{p_s\omega^2}{3\sqrt{2}\pi\omega_m^3}\frac{1}{\sqrt{1-\omega^2/\omega_m^2}}
\end{equation}
where the prefactor $1/3$ results from the usual convention of normalizing the three 
modes per atom to 1. Since we are only interested in the density of states well below 
$\omega_m$, we neglect the square root term. Then $g_s(\omega)/\omega^2$ is simply 
constant
\begin{equation}\label{sdens}
\frac{g^\circ_s(\omega)}{\omega^2}=\frac{p_s}{3\sqrt{2}\pi\omega_m^3}.
\end{equation}
This constant is to be compared with the constant $3/\omega_D^3$, obtained by dividing 
the Debye density of states by $\omega^2$. For the average $Gv/k_BT_g=17.6$ (see Fig. 2),
one finds with Eqs.~(\ref{omd}) and (\ref{ps}) a ratio of 0.6 between the density of 
the soft shear modes and the Debye one. Thus one has a rather large number of soft gliding triangle modes at low frequency, comparable with the number of sound waves.

\subsection{The shape of the boson peak}

The soft shear modes are not exact eigenmodes of the harmonic vibrations of the glass.
They interact with the other vibrations and in particular with the sound waves. The
low frequency shear modes can be understood as the cores of quasi-localized or
resonant vibrations. They are bilinearly coupled to the other modes. This system of
localized soft vibrations coupled bilinearly to extended modes (sound waves) is the
basis of the soft potential model \cite{KKI,BGGS:91,BGGPRS:92}. 

Through their interaction with the sound waves, there is also an interaction between
the soft shear modes. The bilinear interaction of the soft modes with the much larger 
number of higher frequency modes  causes a downshift of the soft modes, some of which
even become harmonically unstable. Stabilization via anharmonicity results in
a soft mode spectrum linear in $\omega$ below a frequency $\omega_c$ given by the
interaction strength, while at low frequency one has a DOS $\propto \omega^4$.
The crossover between the $\omega^4$ and $\omega$ regions then leads to the
boson peak.\cite{GPS:03} The total number of soft modes is not changed by these
frequency shifts. It has been shown that this interaction mechanism can explain
the nearly universal density of observed two level systems in glasses.\cite{PSG:07}

For the purpose of the present paper, it suffices to know that the exact shape of the boson
peak requires a more detailed theoretical treatment. Here, we restrict ourselves to the conclusions
which are easily accessible from our postulate for the eigenvector of the excess modes. As will be seen in the next subsection, these conclusions include a calculation of the damping and softening of the sound waves from the measured spectrum of a given metallic glass. 
 
\subsection{Coupling, damping and softening}

The coupling of a gliding-triangle soft shear mode to an external shear strain is given by the derivative of the elastic energy of eq. (\ref{etot}) with respect to both the external strain $e_m$ and the internal coordinate $e_c$
\begin{equation}\label{detot}
\frac{\partial^2 E_{el}}{\partial e_m\partial e_c}=-6Gv
\end{equation}
if the external shear strain happens to lie exactly in the direction of $e_c$. In the general case, the directional average over the five possible shear strain orientations must naturally be taken into account.

It is usual to express the coupling in terms of a product $\Lambda_tAe_m$ of the shear strain and the soft mode coordinate $A$, defined via eq. (\ref{normal}). From $\Lambda_t$, one can define a frequency $\omega_t$ via
\begin{equation}\label{omtdef}
	\omega_t^2=\frac{\Lambda_t^2}{Mv_t^2}.
\end{equation}
Sorting all the factors out, one finally gets
\begin{equation}\label{omt}
\omega_t=\frac{2^{5/6}\pi^{1/3}}{3^{2/3}}\frac{v_t}{v^{1/3}},
\end{equation}
about a factor of 3.5 lower than the Debye frequency.

With $\omega_t$, the damping $\Gamma$ of the transverse sound waves (the full width at half maximum in a Brillouin experiment) is given by
\begin{equation}\label{gamt}
	\Gamma_t=\frac{3\pi}{2}\omega_t^2 g_s(\omega),
\end{equation}
so the treatment predicts a damping of the transverse sound which increases with the frequency or wavevector squared. The Ioffe-Regel condition $\Gamma=\omega/\pi$ is reached at the frequency $\omega_{JRt}$
\begin{equation}\label{omjrt}
\omega_{JRt}=\frac{2\omega_D^3}{9\pi^2\alpha\omega_t^2}.
\end{equation}
Here $\alpha$ is the ratio between Debye density of states and soft shear wave density of states, which in the average metallic glass should be close to 1. For the average case with $Gv/k_BT_g=17.6$, $\omega_{JRt}$ lies at 0.23 $\omega_D$.

The Ioffe-Regel limit for the longitudinal sound waves lies markedly higher, because the gliding-triangle modes couple only to the shear (or essentially so; we come back to this point in the discussion). Since the longitudinal elastic constant is given by $C_{11}=B+4G/3$, where $B$ is the bulk modulus, a pure shear coupling implies $\Lambda_l^2=4\Lambda_t^2/3$ for the longitudinal coupling constant. With again $\omega_l^2=\Lambda_l^2/Mv_l^2$, $\omega_l^2$ will be a factor of $4v_t^2/3v_l^2$ smaller than $\omega_t^2$, with $v_l/v_t\approx 2.26$ a factor of 0.26 smaller \cite{samwer}, which makes $\omega_{JRl}$ a factor of nearly 4 higher than $\omega_{JRt}$.

The influence of the soft shear modes on the shear modulus is given by
\begin{equation}\label{deltag}
	\frac{\Delta G}{G}=3\omega_t^2\int_0^{\omega_{max}}\frac{g_s(\omega)}{\omega^2}d\omega,
\end{equation}
where $\Delta G$ is the difference between $G$ above and below the boson peak.

\section{Comparison to experiment}

\subsection{Vit-4, a heavily studied metallic glass} 

\begin{figure}[b]
\hspace{-0cm} \vspace{0cm} \epsfig{file=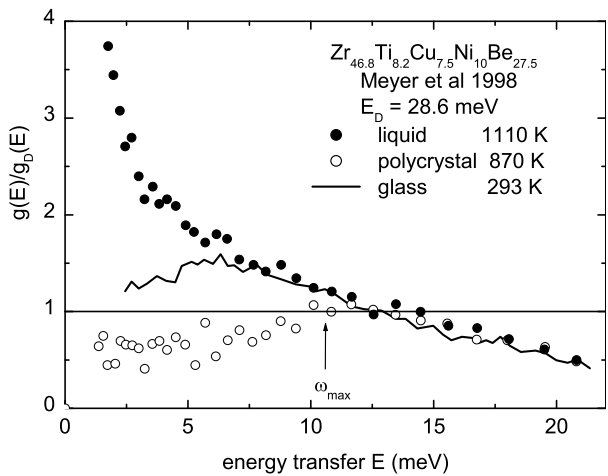,width=8cm,angle=0} \vspace{0cm}\caption{The boson peak of the model glass Vit-4 in comparison to crystal and liquid, determined from inelastic neutron scattering \cite{meyer1,meyer2} in units of the glass Debye-level $3/E_D^3$. The arrow marks the maximum frequency $\omega_{max}$ up to which one expects gliding triangle modes.}
\end{figure}

The best candidate for a check of these ideas is the metallic glass Vit-4, Zr$_{46.8}$Ti$_{8.2}$Cu$_{7.5}$Ni$_{10}$Be$_{27.5}$, a glass with an exceptionally low critical cooling rate \cite{peker}, for which every relevant quantity has been measured \cite{ohsaka,meyer1,meyer2,lind}, though the mechanical \cite{lind}, density \cite{ohsaka} and neutron \cite{meyer1,meyer2} data were all taken for slightly different compositions. At the glass temperature $T_g=615$ K, the shear modulus is 35 GPa, the bulk modulus 112 GPa \cite{lind} and the atomic volume is 0.017 nm$^3$, so $\hbar\omega_D=27.5$ meV. Fig. 4 shows the boson peak at room temperature determined from inelastic neutron scattering data \cite{meyer1,meyer2} in units of the Debye density of states (at room temperature \cite{lind}, $\hbar\omega_D=28.6$ meV).

If one takes the excess over the Debye density of states as gliding-triangle modes, one can calculate the corresponding damping of the sound waves. One finds that the Ioffe-Regel limit is never reached, even for the transverse sound waves. At the boson peak, the damping is about half the Ioffe-Regel limit. Since the longitudinal waves are even less damped, one expects rather sharp longitudinal Brillouin peaks in an inelastic x-ray experiment. This was in fact measured for another metallic glass \cite{nizr}, Ni$_{33}$Zr$_{67}$. In this case, one finds a $Q^2$ dependence above $\omega_{max}$, which probably has nothing to do with the boson peak.

\begin{figure}[b]
\hspace{-0cm} \vspace{0cm} \epsfig{file=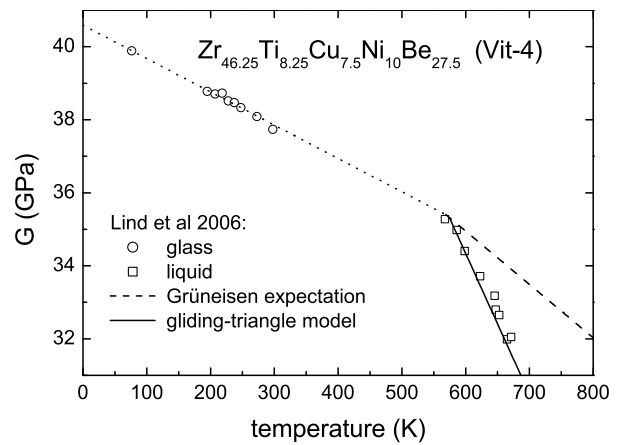,width=8cm,angle=0} \vspace{0cm}\caption{Shear modulus measurement \cite{lind} in glassy and liquid Vit-4. The dotted line corresponds to a Gr\"uneisen parameter $\Gamma_G=6.5$ in the glass.}
\end{figure}

Though the boson peak is small, it is crucial for the mechanical properties and their temperature dependence in the undercooled liquid. One calculates $\Delta G/G=0.08$ from eq. (\ref{deltag}) with the excess modes from Fig. 4. This fits in with the measured difference \cite{lind} of 26 \% of the shear moduli of glass and crystal. The shear modulus Gr\"uneisen parameter $\Gamma_G=\partial \ln{G}/\partial\ln{v}$ is found to be 6.5 from the temperature dependence of $G$ \cite{lind} and the density \cite{ohsaka} in the glass phase. Consequently, the density ratio 1.027 between crystal and glass explains two thirds of the shear modulus difference, the gliding-triangle modes explain the rest.

In the undercooled liquid, the number of gliding-triangle modes increases exponentially according to eq. (\ref{ps}) (replacing $T_g$ by $T$). They do not only increase their number, but also shift to lower frequency, as seen in Fig. 4. However, if one makes a rough estimate by a linear extrapolation of the liquid data in Fig. 4 to the frequency zero, the increase of the area under the curve follows in fact eq. (\ref{ps}). Taking $\Delta G/G=0.08$ as the effect of the gliding-triangle modes frozen in at $T_g$, eq. (\ref{ps}) predicts an additional reduction of $G$ of 0.023 GPa/K in the melt. This is almost exactly found in experiment \cite{lind} (see Fig. 5). The reduction in the liquid is a factor of 2.4 larger than the Gr\"uneisen expectation on the basis of the thermal expansion \cite{ohsaka}. The difference can be frozen into the glassy state by quenching. It is obviously due to configurational changes of the melt. The finding is of great importance for the understanding of liquids. Our model gives a quantitative physical explanation for it: it is due to shear instabilities, stabilized by the surrounding matrix and requiring a formation energy of about three times $k_BT_g$, which reduce the shear modulus.

The experiment \cite{lind} reveals also an additional decrease of the bulk modulus with increasing temperature, though not quite as pronounced as the one of the shear modulus. If this is also attributed to our shear defects, they must couple to the compression. From the data \cite{lind}, the coupling constant ratio $\Lambda_l/\Lambda_t$ of these defects should be 1.69 rather than the value $\sqrt{4/3}$ for a pure shear defect, close to the value 1.6 found for the tunneling states in glasses \cite{berret}. We will come back to this point in the next section.

\subsection{Soft potential model}

The gliding-triangle model makes some rather detailed predictions for the low temperature glass anomalies \cite{phillips}. The easiest way to do this part of the comparison to experiment is to look at the relation between the picture given here and the soft potential model, the standard model in the field for which an extensive comparison with experimental data already exists \cite{parshin,rb,ramos}.

The soft potential model \cite{parshin,rb,ramos} connects the boson peak modes to the tunneling states \cite{phillips} and to low-barrier relaxation. It postulates a constant density of first and second order terms $D_1$ and $D_2$ around the value zero for the potential
\begin{equation}\label{softpot}
	E_{soft}=W(D_1x_s+D_2x_s^2+x_s^4).
\end{equation}
The coordinate $x_s$ relates again to the coordinate $A$ of the soft mode with $E_{kin}=\dot{A}^2/2$. Let us denote the fourth order term in this coordinate by $v_4A^4$. Then the zero-point energy of the purely quartic potential is by definition $W$, given by the equality of kinetic and potential energy at $A_0$
\begin{equation}\label{w}
	W\equiv\frac{\hbar^2}{2A_0^2}=v_4A_0^4.
\end{equation}
This relation defines both $W$ and $A_0$. The kinetic energy results from the uncertainty principle at a confinement within $\pm A_0$. The energy $W$ separates tunneling states below $W$ from vibrational states above. The coordinate $x_s=A/A_0$.

The gliding-triangle modes of the present paper are also distributed around a purely quartic potential (the continuous red line in Fig. 2). Its fourth order term $2\pi^2Gve_c^4$, obtained by the fourth derivation of eq. (\ref{etot}), leads in combination with the definition of $A$ via eq. (\ref{normal}) to
\begin{equation}\label{wsoft}
W=\frac{5^{2/3}\pi^{10/9}}{2^{5/9}3^{20/9}}(Gv)^{1/3}(\hbar^2/Mv^{2/3})^{2/3}.
\end{equation}

Eq. (\ref{wsoft}) supplies an average value $W/k_B$ of 4.4 K for the thirty metallic glasses of the data collection of Johnson and Samwer \cite{samwer}, a value well within the range of those adapted with the soft potential model to a number of nonmetallic glasses \cite{rb}. It combines a very high energy, $Gv$, coming from the fourth order potential term, with the very low kinetic confinement energy $\hbar^2/Mv^{2/3}$. In this context, it is interesting to note that the fourth order term of the quartic potential in Fig. 2 does not result from the short range atomic repulsion, but rather from the decrease of the curvature of the inclusion energy with increasing $e_c$, an anharmonic property of the saddle point of the inclusion energy.

Let us next turn to the tunneling states. Of course, the potential of eq. (\ref{etot}) is in no case a double-well potential with a low barrier able to form a tunneling state, because the negative force constant of the inclusion can at most cancel the positive force constant of the surrounding matrix. In terms of the soft-potential model, the treatment in Section II corresponds to the case where the density $p(D_2)$ is not constant around zero, but rather a $\delta$-function at zero (this explains why one gets an $\omega^2$-rise in the excess density of states rather than the $\omega^4$-rise of the soft potential model). But it was already pointed out in section II that the interaction between the quasilocalized modes is responsible for the shape of the boson peak and the appearance of tunneling states \cite{GPS:03,PSG:07}. However, such a detailed modelling is beyond the scope of the present paper. Here we make no statement on the number of these tunneling states. But we can make a statement on their coupling to the sound waves.

For $W=4.4$ K, a tunneling state with a splitting around 1 K requires $D_2=-6$. The two minima of the double-well potential have a distance $\sqrt{-2D_2}$ in the coordinate $x_s$ of eq. (\ref{softpot}) from each other. Using the definitions of equs. (\ref{w}) and (\ref{normal}), one gets the corresponding distance in $e_c$. Multiplying this distance with the coupling factor $6Gv$ of eq. (\ref{etot}) and again averaging over the five possible shears, one gets the coupling constant $\gamma_t$ of the tunneling state to a transverse sound wave
\begin{equation}\label{gtunnt}
\gamma_t=\frac{2^{10/9}3^{4/9}}{5^{1/3}\pi^{2/9}}\left(\frac{\hbar^2/Mv^{2/3}}{Gv}\right)^{1/6}\sqrt{-D_2}Gv.
\end{equation}
For the data collection on metallic glasses \cite{samwer} this yields an average value of 0.49 eV, not too far away from the average value 0.39 of direct measurements of $\gamma_t$ of the so-called low-temperature "tunneling plateau" in the sound absorption of 18 different inorganic glasses \cite{berret}. In this field, scientists have always been wondering what kind of mode coordinate would be able to give such a strong coupling. Here we have for the first time a detailed answer to this question, at least for an important subgroup of glasses.

The coupling to longitudinal sound waves is described by the analogous coupling constant $\gamma_l$. From the consideration at the end of section II on a pure shear defect like the one proposed here, one would expect a ratio $\gamma_l/\gamma_t=\sqrt{4/3}=1.155$. This does not agree with the experimental finding \cite{berret} $\gamma_l/\gamma_t=1.6$. As pointed out in the previous subsection, the temperature dependence of the elastic moduli in the Vit-4 melt \cite{lind} suggests a factor of 1.69, so metallic glasses in this respect do not differ from the general case. At present, we see no obvious reason for this additional coupling to the compression.

\section{Discussion and summary}

As seen in section III, the Ioffe-Regel limit for longitudinal sound waves for metallic glasses should be by a factor of two to four higher than the one for transverse sound waves. In fact, an inelastic x-ray Brillouin experiment in a metallic glass reveals a relatively low longitudinal sound wave damping \cite{nizr}.

However, the metallic glass result \cite{nizr} is an exception rather than the rule \cite{ruffle}. In the nonmetallic glasses, one usually does find the Ioffe-Regel limit close to the boson peak, maybe slightly above the boson peak. How is this possible?

The question is related to three ratios: The ratio of the boson peak height to the Debye level, the ratio of the coupling constants $\Lambda_l^2/\Lambda_t^2$, which for a pure shear defect is $4/3$, and the ratio $v_l^2/v_t^2$. As it turns out \cite{samwer}, this last ratio is about 5.1 in the metallic glasses, unusually high (compare silica, where this ratio is 2.3). This and the weakness of the boson peak in many metallic glasses are the two reasons for the weak longitudinal damping. A simulation \cite{schober} in a soft sphere glass, where the ratio $v_l^2/v_t^2$ was even higher than for real metallic glasses, showed a very weak damping of the longitudinal waves.

There seems to be some as yet unexplained coupling of the boson peak modes to the compression. As pointed out in section III, the ratio $\Lambda_l^2/\Lambda_t^2$ seems to be $1.6^2=2.56$ rather than $4/3$. This feature is missing in our model. If one ascribes it to different packing factors for octahedron and tetrahedra pair, these should differ by about 18 \%. This explanation seems not to be very convincing. We rather think that the absence of the compression coupling in our model comes from the neglect of the atomic roughness of the shell surrounding the six central atoms, which introduces a random coupling to the compression. 

Despite this small uncertainty, the evidence for the validity of the gliding triangle model in section III shows that it is able to compete with other explanations of the boson peak \cite{schirmacher,elliott,nakayama,gotze,sokolov,parisi,ruocco}. There is no adaptable parameter; one only needs the shear modulus, the atomic mass and the atomic volume. The picture provides a microscopic basis for the empirical soft-potential model \cite{gurevich,parshin,rb,ramos}, not for all glasses, but for the subgroup of metallic glasses.

The question is: To which extent is this answer universal? How should one generalize it to other glasses, glasses with covalent bonds forming a random network, molecular glasses with the additional degree of freedom of the molecular rotation, to the random chain structure of polymers? The answer to the question requires further studies. It might be an intelligent guess, however, to assume the following two general features:

(i) The boson peak and the tunneling modes are always of the same nature, as postulated by the soft potential model

(ii) Boson peak modes and tunneling modes are due to small regions with a strong elastic shear misfit, lying close to a saddle point of the inclusion energy between two stable structural minima of the inclusion.

The validity of these two assumptions would explain the universality of the low temperature anomalies in glasses, including those at slightly higher temperatures where the boson peak modes dominate.

\section{Appendix: The 12 saddle points in cubic notation}

In an {\it fcc} close packed crystal, the six atoms of the octahedron have the coordinates $(\pm a/2,0,0)$, $(0,\pm a/2,0)$ and $(0,0,\pm a/2)$ in units of the lattice constant $a$, related to the atomic volume by $a^3=4v$.

\begin{figure}[b]
\hspace{-0cm} \vspace{0cm} \epsfig{file=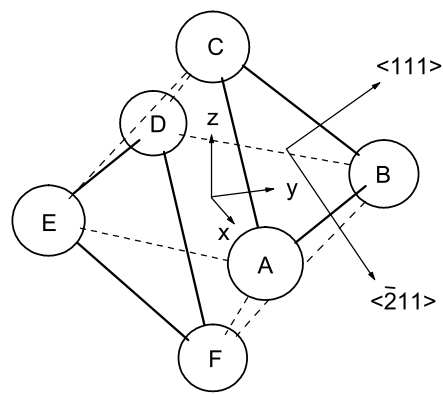,width=8cm,angle=0} \vspace{0cm}\caption{The octahedron and one of its twelve gliding triangle modes in cubic notation. Triangle ABC glides in the $<\bar{2}11>$ direction, triangle DEF in the opposite direction.}
\end{figure}

Let us first consider the gliding-triangle motion of the two triangles in the $(111)$-plane along $<\bar{2}11>$ in Fig. 6. The upper triangle ABC moves opposite to the lower triangle DEF. This gliding motion involves not only a shear, but also a rotation. To get the rotation-free shear, let us define the two unit vectors $\vec{a}_1=<\bar{2}11>/\sqrt{6}$ and $\vec{a}_2=<111>/\sqrt{3}$. With the origin of the position vector $\vec{r}$ at the center of the octahedron, the displacement vector $\vec{u}(\vec{r})$ of the corresponding pure shear is given by
\begin{equation}\label{uvr}
	\frac{2}{e_c}\vec{u}=\left(\vec{r}\vec{a}_2\right)\vec{a}_1+\left(\vec{r}\vec{a}_1\right)\vec{a}_2,
\end{equation}
where $e_c$ is the shear angle of the gliding triangle mode of Fig. 6.

From eq. (\ref{uvr}), one can calculate the shear. In Kittel's notation \cite{kittel} (diagonal elements $e_{xx}=\partial u_x/\partial x$, nondiagonal elements $e_{xy}=\partial u_x/\partial y+\partial u_y/\partial x$), one finds for the gliding triangle mode of Fig. 6 in units of $e_c\sqrt{2}/6$ the diagonal elements $e_{xx}=-4$, $e_{yy}=2$ and $e_{zz}=2$, the nondiagonal elements $e_{xy}=-1$, $e_{xz}=-1$ and $e_{yz}=2$. The elastic energy of the octahedron is
\begin{equation}\label{eeloct}
	E_{el}=(C_{11}-C_{12}+C_{44})ve_c^2=(2C'+C_{44})ve_c^2,
\end{equation}
where $C'=(C_{11}-C_{12})/2$ is one of the two shear moduli of the cubic crystal and $C_{44}$ is the other. Equating both of them to the isotropic shear modulus $G$ of the glass, we are back to the curvature of eq. (\ref{cos}) at the octahedral site.

\begin{figure}[b]
\hspace{-0cm} \vspace{0cm} \epsfig{file=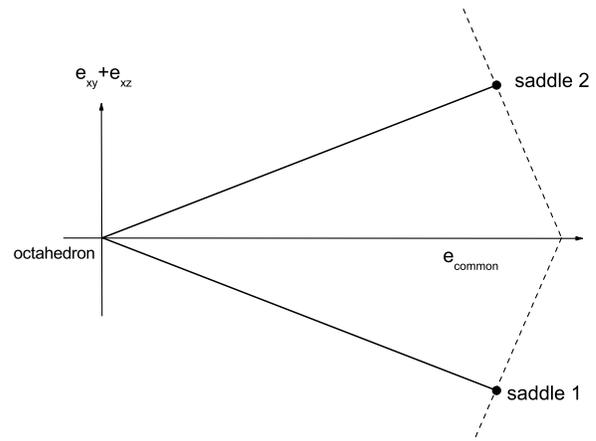,width=8cm,angle=0} \vspace{0cm}\caption{The two neighboring saddle points with a strong common shear $e_{common}$ and a weak opposite shear $e_{xy}+e_{xz}$ with respect to the undistorted octahedron.}
\end{figure}

Two thirds of the elastic energy of eq. (\ref{eeloct}) come from the diagonal elements, a dilatation in the $y$ and in the $z$-direction accompanied by a factor of two larger contraction in the $x$-direction. As it turns out, this special shear is a common feature of four of the twelve saddle point distortions, each coming from a different triangle plane. They only differ in the nondiagonal elements, the one from the $(1\bar{1}1)$-plane having $e_{xy}=1$, $e_{xz}=-1$ and $e_{yz}=-2$, the one from the $(11\bar{1})$-plane having $e_{xy}=-1$, $e_{xz}=1$ and $e_{yz}=-2$ and finally the one from the $(\bar{1}11)$-plane having $e_{xy}=1$, $e_{xz}=1$ and $e_{yz}=2$. This last one is closest to the first one, because the shear occurs in the same shear plane, with only a slight rotation of the shear directions. In the five-dimensional shear space, the lines connecting those two saddle points with the octahedron have an angle of only 38.9 degrees (Fig. 7).

The closeness has consequences for the probability density $p_s$ of section II. This was calculated integrating the partition function $Z$ only over the connection line between octahedron and saddle point, and then multiplying with the number of saddle points. In principle, one needs the full integration over the five-dimensional shear space. As long as the saddle points are well separated in phase space, the integration over the irrelevant degrees of freedom can be omitted \cite{vineyard}. But if two saddle points are as close together as in Fig. 7, the integration over their neighborhood (along the dashed lines in Fig. 7) leads to double-counting. In a crude approximation, we counted the four saddle points with the common diagonal shear elements as only two in eq. (\ref{ps}) (remember that the equation is only needed to demonstrate that the density of soft gliding triangle modes is comparable to the one of the sound waves).

\end{document}